# Enhancing the prediction of publications' long-term impact using early citations, readerships, and non-scientific factors

Giovanni Abramo (*corresponding author*)
  Laboratory for Studies in Research Evaluation
  Department of Engineering and Science
  Universitas Mercatorum
  Piazza Mattei, 10 - 00186 Rome, Italy
  giovanni.abramo@unimercatorum.it
  ORCID: 0000-0003-0731-3635

Tindaro Cicero
  Laboratory for Studies in Research Evaluation
  Department of Engineering and Science
  Universitas Mercatorum
  Piazza Mattei, 10 - 00186 Rome, Italy
  tindaro.cicero@unimercatorum.it
  ORCID: 0000-0003-1926-8317

Ciriaco Andrea D'Angelo
  Dipartimento di Ingegneria dell'Impresa
  Università degli Studi di Roma "Tor Vergata"
  Via del Politecnico, 1 - 00133 Rome, ITALY
  dangelo@dii.uniroma2.it
  ORCID: 0000-0002-6977-6611

**Abstract**
This study aims to improve the accuracy of long-term citation impact prediction by integrating early citation counts, Mendeley readership, and various non-scientific factors, such as journal impact factor, authorship and reference list characteristics, funding and open-access status. Traditional citation-based models often fall short by relying solely on early citations, which may not capture broader indicators of a publication's potential influence. By incorporating non-scientific predictors, this model provides a more nuanced and comprehensive framework that outperforms existing models in predicting long-term impact. Using a dataset of Italian-authored publications from the Web of Science, regression models were developed to evaluate the impact of these predictors over time. Results indicate that early citations and Mendeley readership are significant predictors of long-term impact, with additional contributions from factors like authorship diversity and journal impact factor. The study finds that open-access status and funding have diminishing predictive power over time, suggesting their influence is primarily short-term. This model benefits various stakeholders, including funders and policymakers, by offering timely and more accurate assessments of emerging research. Future research could extend this model by incorporating broader altmetrics and expanding its application to other disciplines and regions. The study concludes that integrating non-citation-based factors with early citations captures a more complex view of scholarly impact, aligning better with real-world research influence.

**Acknowledgment**
We wish to thank Elsevier for providing us with Mendeley's readership data.

**Highlights**

- We integrate non-citation factors with early citations to predict long-term research impact
- Early citations and Mendeley readership are key predictors of citation longevity
- Open-access status and funding boost short-term impact but diminish over time
- Our model aids funders and policymakers in assessing research influence quickly and accurately
- Results validate a comprehensive approach to forecasting a publication's future relevance

# 1. Introduction

Accurately assessing the long-term impact of scientific publications is essential for understanding knowledge development, evaluating research contributions, guiding funding allocation, and shaping scientific policies and strategies. Traditionally, citation counts have been the primary measure of a publication's influence (Abramo, 2018; Bornmann & Daniel, 2008), with early citations often serving as a key indicator of future impact. As citations accumulate over time, they collectively represent the scholarly influence of a publication throughout its life cycle[1] (from date of publication to date of last citation), which can span decades.

However, waiting for decades to fully assess the impact of a publication is impractical for decision-makers who rely on timely evaluations to guide research investments and policy decisions. Although extending the citation-time window improves the accuracy of impact assessments, it comes at the cost of delayed insight. As a result, there is a natural tradeoff between timeliness and accuracy. This challenge calls for bibliometricians to improve the predictive power of early citation data, allowing for more immediate yet reliable assessments of long-term impact.

Scientific impact is a complex and dynamic process influenced by many factors beyond just early citations. Recent advances in bibliometrics and data analytics have opened new avenues for refining models that forecast the long-term influence of publications. Various methods have been proposed to enhance early citation predictions, falling broadly into four categories: mathematical statistics, traditional machine learning, deep learning, and graph theory (Xia et al., 2023). The forecasting approach introduced in this study aligns with the first category.

In addition to early citation trends, non-scientific factors such as the journal's prestige, authors' institutional affiliations, collaboration networks, and media coverage have gained recognition as important drivers of a publication's visibility and impact over time (Mammola, Piano, Doretto, Caprio, & Chamberlain , 2022; Xie et al., 2019; Tahamtan, Safipour Afshar & Ahamdzadeh, 2016). We aim to develop a more accurate and robust predictive model by integrating these non-scientific factors with early citation data. This enhanced model will provide valuable tools for a wide range of stakeholders—including researchers, funders, and policymakers—enabling them to make more informed decisions about emerging research and to allocate resources more effectively.

We hypothesize that incorporating these diverse predictors will yield more accurate forecasts than models based solely on early citations. Our approach aims to provide a deeper understanding of the multifaceted drivers behind research impact and establish a comprehensive framework for evaluating how scientific work garners influence throughout its life cycle. Moreover, this framework offers practitioners a more precise tool for assessing the influence of scientific publications.

In particular, we try to answer the following research questions:
- What role do features beyond early citations play in predicting a publication's long-term impact?
- How does the influence of these predictors change with the length of the citation time window?

---

[1] The life cycle of a research product refers to the stages it undergoes in terms of citations, starting from its publication and continuing until its eventual decline, when it is no longer cited.

- How much does the predictive power of early citations improve when these additional predictors are also considered?

To address these questions, we will employ an OLS model, using the scholarly impact of Italian publications from 2010–2012 in the Web of Science (WoS) as the response variable, represented by citations received 11 years post-publication. The independent variables will include short-term impact, measured by citations within varying time windows of 0 to 6 years, readership metrics, and various non-scientific characteristics related to the publication's authorship, content, and venue.

In the following section, we review the relevant literature on late citation forecasting and the role of non-scientific factors. Section 3 outlines the field of observation, data sources, and methods employed. Results are presented in Section 4, followed by discussion, conclusions, and suggestions for future research in Section 5.

## 2. Review of the literature

Predicting the long-term impact of scientific publications is a very active area of research, driven by the increasing volume of scientific output and the need to evaluate its research influence. The breadth of the literature is so extensive that any attempt to synthesize it in a two-page review is bound to overlook some important contributions to the field. For a more comprehensive and in-depth analysis, readers are encouraged to consult: i) Hou et al. (2019), who reviewed various methods and applications for predicting publication impact; ii) Bai et al. (2017), who focused specifically on methodologies for predicting the impact of scholarly articles; and iii) Xia, Li, and Li, (2023) who offered a taxonomy of approaches. Additionally, they updated and expanded previous surveys by incorporating the latest advances in machine learning, including deep learning, recurrent neural networks (RNNs), convolutional neural networks (CNNs), and graph neural networks (GNNs).

Several methodologies and factors contribute to understanding and predicting the future impact of a publication, typically measured by citation counts or other indicators of scholarly recognition. Below is a review of key concepts, models, and approaches used in the literature to predict the long-term impact of scientific publications.

It is well known that different types of publications (such as articles, reviews, and conference proceedings) show varying citation patterns. Reviews, for example, are typically cited more frequently than original research articles, while conference proceedings tend to have lower citation rates (Aksnes, 2006; Moed, 2005; Narin & Hamilton, 1996).

Citation-based models focus on understanding and predicting the long-term citation trajectory of a publication. Studies have consistently shown that early citations are strong predictors of future impact. Papers that garner significant attention in the first few years tend to maintain a higher citation rate over the long term. Bornmann and Daniel (2010) found that papers with high early citation rates often sustain their trajectory, although the slope of citation growth may vary across disciplines.

Time series models have been used to predict long-term citation dynamics by treating citations as sequential events. Wang, Song, and Barabási (2013) proposed a model based on the "ageing" of papers, suggesting that citation rates peak early in a paper's life cycle and gradually decay as the paper becomes outdated or overshadowed by newer research.

These models are often represented through exponential decay functions or other statistical models that capture the half-life of citations.

Survival analysis, initially used in fields like medicine, has been adapted to model the "survival" of citations over time, estimating how long a paper will continue to receive citations. Parolo et al. (2015) and Eom and Fortunato (2011) applied survival models to identify citation patterns that lead to sustained impact.

The "rich get richer" principle, also known as preferential attachment or cumulative advantage, is commonly applied in citation prediction. This model posits that papers with a high initial number of citations are more likely to receive additional citations. Price (1976) was one of the first to introduce this idea, which has been empirically validated in various studies (Newman, 2018).

Research has also focused on the content characteristics of papers that contribute to long-term impact, such as novelty, interdisciplinarity, and the concept of "disruptive" science. Uzzi, Stringer, and Jones (2013) found that papers combining novel ideas with conventional knowledge tend to have a higher long-term impact. Their model shows that a balance between novelty and familiarity is key to predicting which papers will receive sustained attention. Papers that span multiple fields are often cited across different domains, leading to a more sustained and broader citation impact (Abramo et al., 2017). Porter and Rafols (2009) found that interdisciplinary papers tend to experience delayed citation growth but often achieve greater long-term influence. Foster, Rzhetsky, and Evans (2015) introduced a measure of how "disruptive" a paper is, based on whether it builds on existing knowledge or opens up new research pathways. Disruptive papers tend to be cited differently from incremental ones, and their citation patterns are harder to predict using traditional models (Leahey, Lee, & Funk, 2023).

Altmetrics, such as mentions on social media platforms (e.g., Twitter), blog posts, or references in policy documents, have emerged as complementary to traditional citation counts (Shema, Bar-Ilan, & Thelwall, 2014; Sud & Thelwall, 2014)). Research has explored whether early altmetric attention can predict future citation impact. Eysenbach (2011) found that tweets within the first few days of publication were associated with higher citation rates over time, suggesting that social media attention could be an early indicator of scientific impact. However, the relationship between altmetrics and long-term impact remains complex, as not all fields or types of papers are equally represented on social platforms.

Some models combine traditional citation data with altmetrics to improve prediction accuracy. Thelwall and Nevill (2018) found that combining different types of early metrics can provide a more nuanced prediction of long-term impact, particularly for interdisciplinary research that may not immediately accumulate citations but garners attention through non-academic channels.

Recent developments in machine learning have enabled more sophisticated approaches to predicting the long-term impact of scientific publications (Himani et al., 2022). These models often incorporate various features beyond citation data, including the paper's content, author characteristics, and network effects (Qiu & Han, 2024; Abrishami & Aliakbary, 2019).

Natural language processing (NLP) techniques are used to analyze the content of papers to predict impact. Factors such as the novelty of the research, its interdisciplinarity, and the use of specific keywords or phrases have been linked to higher citation potential. Yau et al. (2014) applied topic modelling and sentiment analysis to extract latent factors correlating with long-term citation trends.

Several studies incorporate author-level and network-based features to improve prediction accuracy. Sarigöl et al. (2014) demonstrated that author collaboration networks, particularly those involving central or highly cited authors, are strong predictors of future paper citations. Similarly, papers authored by well-established or highly productive researchers tend to have a greater long-term impact (Petersen et al., 2014).

The venue of publication can also be an important predictor of long-term impact. Papers published in high-impact journals tend to accumulate more citations, and models that incorporate journal impact factors often show improved prediction performance (Abramo, D'Angelo, and Felici, 2019; Levitt & Thelwall, 2011). However, Larivière and Gingras (2010) suggest that the relationship between journal impact factor and individual paper impact is not always linear, leading to nuanced adjustments in prediction models.

Recently, artificial intelligence methods have been used to forecast the future influence of publications (Thelwall, 2024; Koshua & Thelwall, 2024; Cárdenas, 2023). However, research on this topic has been limited to individual disciplines.

The citation patterns and long-term impact of publications can vary significantly across disciplines. Even within the same discipline, the citation dynamics of papers can vary considerably (Garfield, 2006; Wang et al., 2021; Yan & Ding, 2010). Several studies have explored these differences and incorporated field-specific factors into prediction models (Xia, Li, & Li, 2023; Abramo, D'Angelo, Felici, 2019). Some disciplines, such as biomedical research, exhibit rapid citation accumulation, while others, like the humanities or mathematics, may have longer citation half-lives. Researchers have adjusted prediction models to account for these field-specific dynamics. For instance, Wang, Song, and Barabási (2013) found that in fast-moving fields, early citation indicators were more reliable. In contrast, longer observation windows were necessary in slower fields to make accurate predictions.

Several studies have compared different models to predict long-term scientific impacts, such as citation-based models, network models, and machine learning approaches. These studies often aim to determine the most effective model for capturing the dynamics contributing to research papers' sustained influence over time.

Penner et al. (2013) compared the ability to forecast future citations of several models, including regression-based models, cumulative advantage (preferential attachment), and network-based models. They found that the initial citation rate is a strong predictor but that more complex network-based models can improve prediction accuracy.

Acuna, Allesina, and Kording (2012) compared simple citation-based models with more complex machine learning models to predict the future impact of researchers. The authors used linear regression, random forests, and support vector machines (SVM) to model citation dynamics and showed that models incorporating early career citation rates outperformed models based solely on cumulative citation counts.

Mazloumian et al. (2011) compared citation-based models, growth models, and models incorporating author reputation and co-authorship networks to predict long-term scientific success. They introduced the concept of "citation boosts" (sudden increases in citations) and found that models considering boosts alongside traditional citation counts provided more accurate long-term predictions.

Kosteas (2020) systematically compared multiple models, including linear regression, logistic regression, and machine learning methods like random forests. He explored the predictive power of early citations (within the first two years) and other factors, such as the paper's field and journal impact factor, concluding that machine learning models generally outperform simpler regression models in forecasting long-term impact.

Yu, Wang, and Guo (2014) compared traditional statistical models, such as autoregressive models, with machine learning algorithms like SVM and decision trees to predict future citation impact. The authors demonstrated that machine learning methods, particularly those incorporating time series segmentation, can significantly improve prediction accuracy over traditional citation-based models.

Li et al. (2019) compared deep learning models (deep neural networks) with more traditional citation-based models, such as linear and logistic regression. The study found that deep learning models could capture complex non-linear patterns in citation trajectories, outperforming simpler models in prediction tasks.

Recently, Zhang and Wu (2024) introduced an innovative prediction methodology that employs multiple models, each customized for specific research domains, while incorporating early citation data. Utilizing these domain-specific models and early citation information, their approach outperforms four leading baseline methods in most cases, significantly improving the accuracy of citation impact predictions across various academic papers.

Our study proposes a forecasting approach grounded in mathematical statistics to better predict the impact of publications. Prior research has identified an association between various non-scientific factors and the influence of publications, and we aim to enhance prediction accuracy and robustness by integrating these non-scientific factors with early citation counts.

The type of access a publication has—whether open access (OA) or non-open access (non-OA)—affects knowledge transferability. Numerous studies have demonstrated that OA articles tend to attract significantly more citations than their non-OA counterparts (Yu et al., 2022; Wang et al., 2015; Gargouri et al., 2010). Likewise, longer articles are generally more cited than shorter ones (Xie et al., 2019; Elgendi, 2019; Fox, Paine, & Sauterey, 2016; Ball, 2008). Linguistic features of the title and abstract, too, appear to influence citation rates (Heßler & Ziegler, 2022; Rossi & Brand, 2020; Abramo, D'Angelo, & Di Costa, 2016; Stremersch et al., 2015; Didegah & Thelwall, 2013). Furthermore, the degree of a publication's interdisciplinarity correlates with its citation impact (Chen, Arsenault, & Larivière, 2015; Yegros-Yegros, Rafols, & D'Este, 2015; Abramo, D'Angelo, & Di Costa, 2017).

Other factors, such as authorship features, also affect citation rates. Studies show that the number of authors is positively associated with citation counts (Talaat & Gamel, 2022; Fox, Paine, & Sauterey, 2016; Abramo & D'Angelo, 2015; Didegah & Thelwall, 2013; Wuchty, Jones, & Uzzi, 2007), as are the number of affiliations and countries represented (Sanfilippo, Hewitt, & Mackey, 2018; Narin & Whitlow, 1990; Glänzel & De Lange, 2002). Publications authored by native English speakers also tend to receive more citations (Hyland, 2016; Ammon, 2012; Meneghini & Packer, 2007; Van Leeuwen & Moed, 2002).

The characteristics of the reference list can further influence citation impact. Longer reference lists have been found to correlate with increased citation counts, especially when they include highly cited and recent publications (Fox, Paine, & Sauterey, 2016; Alimohammadi & Sajjadi, 2009; Sivadas & Johnson, 2015; Jiang, He, & Ni, 2013; Liu et al., 2022; Mammola et al., 2021). Additional factors include the diversity of cited fields and the cognitive distance of these citations (Wang, Thijs, & Glänzel, 2015), as well as the proportion of self-citations (Ruan et al., 2020).

Exogenous factors, such as the prestige of the journal (Traag, 2021; Mammola et al., 2021) and the extent of social media exposure (Özkent, 2022), also contribute to citation advantages. Mendeley readership, in particular, has been identified as a significant predictor of future citation counts (Zahedi, Costas, & Wouters, 2014; Mohammadi & Thelwall, 2014; Haustein et al., 2014). Moreover, studies indicate that financial support often correlates with higher citation rates (Wang, Veugelers, & Stephan, 2017; Rigby, 2013; Boyack & Jordan, 2011; Aksnes, 2003; Lewison & Dawson, 1998).

While many non-quality-related factors influence citation counts, not all can be consistently measured on a large scale.

## 3. Data and methods

Our analysis focuses on 2010–2012 Web of Science (WoS) publications in which at least one author lists "Italy" as their affiliation country, with document types limited to articles, reviews, and conference proceedings. These publications were sourced from the National Citation Report-Italy through a licensing agreement with Clarivate Analytics. Access to this report enables us to obtain predictors such as citation counts at various intervals post-publication, financial support information, access type, number of references, etc. The dataset includes 293,538 publications in all. Through the WoS classification schema, each of them is assigned to the subject category (SC) of the hosting journal.[2] For each publication, we extract information regarding various features, including citation impact, characteristics of the byline, attributes related to the content and publication venue, and features associated with the publication's bibliography.

We use an OLS model to examine the role of features beyond early citations in predicting a publication's long-term impact. The normalized impact (IMPACT_$t_i$) is measured by normalized citations, defined as the citation count divided by the average citations of all Web of Science (WoS) publications within the same subject category (SC) and publication year. Citations assessed 11 years post-publication serve as our response variable for long-term impact. Short-term impact is calculated over varying citation windows from 0 to 6 years. For each of these time windows, we also compute normalized Mendeley readership counts (READ_$t_i$), scaled to the average readership of WoS publications in the same SC and publication year.

Additional features related to the publication's content and venue include:
- Authorship: Number of authors (AUTH) and a dummy variable for native English-speaking authors (ENG) to account for potential linguistic advantage.
- International collaboration: A dummy variable for publications with multiple countries listed in the affiliations (FOREIGN).
- Funding: A dummy variable for funded research, based on acknowledgments (FUNDING).
- Open access status: A dummy variable for Green, Hybrid, and Gold open access publications (OPEN).
- Length: Number of pages (PAGES).
- References: Reference list length as the count of cited references (REFER).
- Journal impact: Normalized impact factor of the source journal in the publication year (IF).

---
[2] Publications hosted in multicategory journals are fully assigned to all SCs of the journal.

- Disciplinary classification: 247 dummy variables (D_SUBCATj) representing the WoS SCs in the sample.
- Document type: Two dummy variables distinguishing articles (ART) and reviews (REWs) from the baseline, conference proceedings.

The list of variables included in the dataset is summarized in Table 1.

*Table 1 – Variable description and related acronym*

| Variables Description | Variable type | Acronym |
|---|---|---|
| Publications' impact after $i$ years | Continuous | IMPACT_t$i$ |
| Mendeley readerships after $i$ years | Continuous | READ_t$i$ |
| Number of authors in the byline | Integer | AUTH |
| Presence of at least one native English-speaking author in the byline | Dummy | D_ENG |
| Two or more countries in the affiliation list | Dummy | D_FOREIGN |
| Publication receiving financial support/funding aknowledgement | Dummy | D_FUNDING |
| Open access publication | Dummy | D_OPEN |
| Publication length (number of pages) | Integer | PAGES |
| Normalized impact factor of the hosting source | Continuous | IF |
| Document type - Article | Dummy | D_ART |
| Document type – Review | Dummy | D_REW |
| Number of references of the publication | Integer | REFER |
| SC of the publication | 247 dummies | D_SUBCAT$_j$ |

As discussed in the upcoming section on descriptive statistics, several variables exhibit high variability relative to the mean, strongly influenced by the presence of outliers. A preliminary analysis using local polynomial smoothing indicates that, for most variables, the relationships appear linear on a logarithmic scale. Consequently, we applied a base-10 logarithmic transformation to integer and continuous variables, adding a unit to prevent data loss due to zero values. This approach aligns with prior literature recommendations for variables such as IMPACT_t$i$ (Bornmann & Leydesdorff, 2017), READ (Costas, Perianes-Rodríguez, & Ruiz-Castillo, 2017), AUTH (De Santis Puzzonia, Beltram, Cicero, & Malgarini, 2018), PAGES, IF, and REFER (Nicolaisen & Frandsen, 2021).

To estimate long-term impact—specifically, 11 years post-publication—as a function of available features, we implemented a series of regression models, summarized in Table 2. This table includes 21 regression models divided into three groups: "Full," "Reduced," and "Completely Reduced" models, each comprising seven iterations.

In the Full models, the dependent variable is the 11-year impact (IMPACT_t11), with a set of independent variables that combines time-varying and time-invariant factors. The time-varying vector X, which includes IMPACT and READ, is adjusted in each iteration. For example, in the first Full model, we use IMPACT_t0 and READ_t0; in the second, IMPACT_t1 and READ_t1; continuing up to IMPACT_t6 and READ_t6 in the seventh model. Other regressors—Z (non-bynary variables such as AUTH, REFER, IF, PAGES), D (binary variables such as D_ART, D_ENG, D_FOREIGN, D_REW, D_FUNDING, D_OPEN), and S (subject categories)—remain constant across iterations.

The same dependent variable (IMPACT_t11) is used in the Reduced models but with fewer regressors. Here, only the variable IMPACT is updated across iterations, with IMPACT_t0 used in the first model, IMPACT_t1 in the second, and so on up to IMPACT_t6. The only time-invariant variable, IF, along with control variables S (subject categories), remains constant throughout. These models examine the effects of early citations and journal impact factor alone, facilitating comparison with similar analyses in the literature (Abramo, D'Angelo & Felici, 2019) and comparing it with the Full model highlights differences in predictive power.

The final set of models, the Completely Reduced models, streamlines the regressor set further. Here, only the impact variable (IMPACT_t) and control variables S (subject categories) are included, with the aim of identifying their effects. This simplified approach is widely used due to the accessibility of early citation data, and comparing it with the Full model highlights differences in predictive power.

*Table 2 – Models estimated and their iterations*

| Model | Equations | Iterations |
|---|---|---|
| Full | $Y_{i(11)} = \beta_1 X_{i,t} + \beta_2 Z_i + \beta_3 D_i + \beta_4 S_i + e_{i(11)}$<br>Where:<br>• Y = impact value of publication *i* at time 11 (IMPACT_t11);<br>• $\beta_1$ = vector of regression coefficients for the vector of variables that are replaced in each model (specifically composed of the variables IMPACT_t and READ_t);<br>• *t* is the time, with $0 \leq t \leq 6$;<br>• $\beta_2$ = vector of regression coefficients for the vector of variables Z that do not vary over time (e.g. AUTH, REFER, IF, PAGES);<br>• $\beta_3$ = vector of regression coefficients for the vector of binary variables D that do not vary over time (e.g., D_ART, D_REW, D_ENG, D_FOREIGN, D_FUNDING, D_OPEN);<br>• $\beta_4$ = vector of regression coefficients for the vector of control variables *S*, representing the SC in which each publication *i* is classified;<br>• $e_{i(11)}$ = error term for publication *i* at time t=11. | 7 iterations ($t_0$–$t_6$): Regressors characterized by Z, D, and S remain constant, but the two variables X are replaced in each iteration (i.e. the vector X is characterized by IMPACT and READ. In the first model, we include IMPACT_t0 and READ_t0; in the second model, IMPACT_t1 and READ_t1 are introduced, and so on, up to IMPACT_t6 and READ_t6 in the seventh model.) |
| Reduced | $Y_{i(11)} = \beta_1 X_{i,t} + \beta_2 IF_i + \beta_3 S_i + e_{i(11)}$<br>Where:<br>• Y = impact value of publication *i* at time 11 (IMPACT_t11);<br>• $\beta_1$ = regression coefficient for the variable that is replaced in each model (specifically IMPACT_t)<br>• *t* is the time, with $0 \leq t \leq 6$;<br>• $\beta_2 IF_i$ = regression coefficient for the variable IF that does not vary over time<br>• $\beta_3$ = vector of regression coefficients for the vector of control variables *S*, representing the SC in which each publication *i* is classified;<br>• $e_{i(11)}$ = error term for publication *i* at time t=11. | 7 iterations ($t_0$–$t_6$): Regressors characterized by IF and S remain constant, but the one variable X is replaced in each iteration (i.e. X is characterized only by IMPACT. In the first model, we include IMPACT_t0; in the second model, IMPACT_t1 and so on, up to IMPACT_t6 in the seventh model.) |
| Completely Reduced | $Y_{i(11)} = \beta_1 X_{i,t} + \beta_2 S_i + e_{i(11)}$<br>Where:<br>• Y = impact value of publication I at time 11 (IMPACT_t11);<br>• $\beta_1$ = regression coefficient for the variable that is replaced in each model (specifically IMPACT_t0)<br>• *t* is the time, with $0 \leq t \leq 6$;<br>• $\beta_2$ = vector of regression coefficients for the vector of control variables *S*, representing the SC in which each publication *i* is classified;<br>• $e_{i(11)}$ = error term for publication *i* at time t=11. | 7 iterations ($t_0$–$t_6$): Regressors characterized by S remain constant, but the one variable X is replaced in each iteration (i.e. X is characterized only by IMPACT. In the first model, we include IMPACT_t0; in the second model, IMPACT_t1 and so on, up to IMPACT_t6 in the seventh model.) |

## 4. Results

### 4.1 Descriptive statistics and correlation analysis

Table 3 presents the descriptive statistics of the variables used in this study, providing an overview of their distribution and key characteristics. The mean values of dummy variables represent the percentage of publications characterized by the specific attribute. Accordingly, 21.9% of publications have at least one native English-speaking author; in 44% of cases, the address list contains more than one country; 33.7% of publications are Open Access; and 49% have received financial funding. Most of the publications are classified as articles (83.8%), while the remaining publications are reviews (6.1%) and conference proceedings (10.1%), which are the residual category not included in the dummy variables.

*Table 3 - Descriptive statistics of analyzed variables (293,538 observations)*

| Variables | Mean | Std Dev. | Min | Max |
|---|---|---|---|---|
| IMPACT_t11 | 1.057 | 2.553 | 0.000 | 309.625 |
| IMPACT_t6 | 0.936 | 1.980 | 0.000 | 196.955 |
| IMPACT_t5 | 0.928 | 1.902 | 0.000 | 193.313 |
| IMPACT_t4 | 0.917 | 1.837 | 0.000 | 210.344 |
| IMPACT_t3 | 0.897 | 1.773 | 0.000 | 227.133 |
| IMPACT_t2 | 0.855 | 1.704 | 0.000 | 239.548 |
| IMPACT_t1 | 0.730 | 1.477 | 0.000 | 193.558 |
| IMPACT_t0 | 0.323 | 0.879 | 0.000 | 78.314 |
| READ_t6 | 1.664 | 4.220 | 0.000 | 951.598 |
| READ_t5 | 1.699 | 4.662 | 0.000 | 1137.025 |
| READ_t4 | 1.722 | 4.957 | 0.000 | 1190.991 |
| READ_t3 | 1.701 | 4.859 | 0.000 | 1077.796 |
| READ_t2 | 1.592 | 4.485 | 0.000 | 967.132 |
| READ_t1 | 1.307 | 3.071 | 0.000 | 487.578 |
| READ_t0 | 0.719 | 1.602 | 0.000 | 216.927 |
| AUTH | 13.849 | 127.614 | 1.000 | 3221.000 |
| D_ENG | 0.219 | 0.413 | 0.000 | 1.000 |
| D_FOREIGN | 0.440 | 0.496 | 0.000 | 1.000 |
| D_FUNDING | 0.490 | 0.500 | 0.000 | 1.000 |
| D_OPEN | 0.337 | 0.473 | 0.000 | 1.000 |
| PAGES | 9.866 | 8.176 | 1.000 | 1504.000 |
| IF | 1.262 | 1.139 | 0.000 | 21.070 |
| D_ART | 0.838 | 0.368 | 0.000 | 1.000 |
| D_REW | 0.061 | 0.239 | 0.000 | 1.000 |
| REFER | 38.465 | 36.365 | 0.000 | 5340.000 |

The 293,538 publications are classified into 248 distinct SCs.[3] These categories are highly heterogeneous, ranging from those with as few as two publications to a maximum of 7,448, with an average of 1,183.6 observations. The Herfindahl-Hirschman Index, which measures the concentration of publications within a single category, is 0.011. This indicates a highly fragmented corpus of publications, with no single SC holding a significant share of the total publications. These variables will be used as control variables in the subsequent analysis, to account for the variability of citation patterns across SCs.

---

[3] Compared to the 254 total SCs in the WoS schema, during the observed period (2010-2012), there are no publications by Italian authors in 6 Art & Humanities SCs (Dance; Folklore; Literature, African, Australian, Canadian; Literature, German, Dutch, Scandinavian; Literature, Slavic; Poetry).

As previously indicated, the available impact values correspond to distinct and specific time points, allowing us to differentiate between short-term impact (within 6 years of publication) and long-term impact (11 years after publication). As might be expected, these impact values are highly correlated, as demonstrated in Figure 1 (left panel). Therefore, we exclude from the outset any predictive model that considers all impact variables simultaneously, as such a model would be vulnerable to severe multicollinearity issues, which would compromise the accuracy of the predictive estimates.

As observed, the correlation between impact values at successive years is very high, with an increasing trend as the years progress. Specifically, the correlation between the impact in the publication year (IMPACT_t0) and the impact in the following year (IMPACT_t1) is 0.77. This value rises to 0.96 between t1 and t2, 0.98 between t2 and t3, and stabilizes at 0.99 for subsequent time points up to 6 years.

In contrast, the long-term impact, measured 11 years after publication, shows a much weaker correlation with the impact in the publication year (0.39). However, the correlation between the long-term impact and the impact two years after publication increases to 0.72.

*Figure 1 - Autocorrelation matrix of impact variables (left panel) and readership (right panel), at time t*

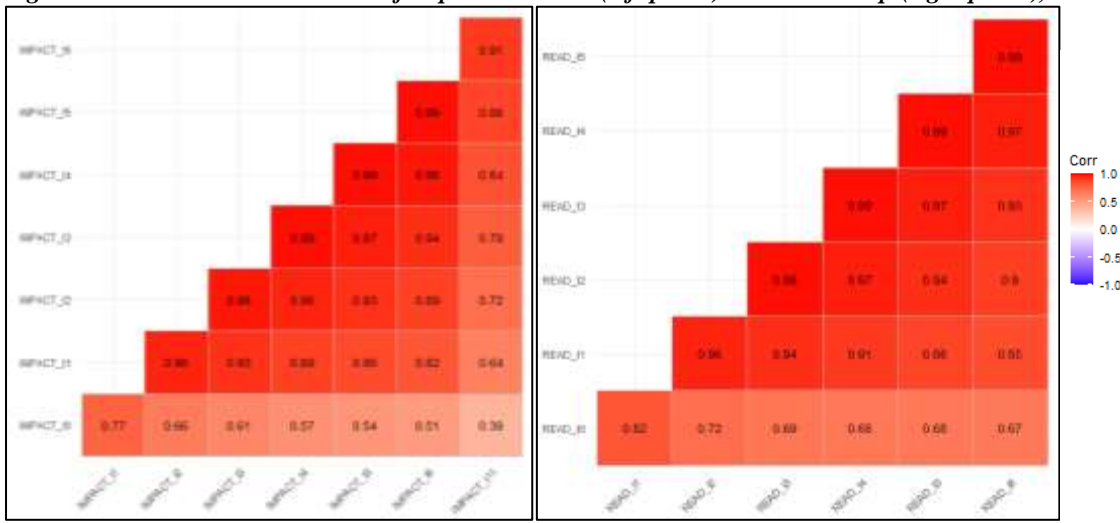

Similar considerations can be made when analyzing the correlation of readership measured at different time points (Figure 1 – right panel). There is consistently a strong correlation between consecutive time points, and the correlation decreases as *t* increases: readership at t=0 (READ_t0) is correlated with readership at t=1 with a value of 0.82. As we fix READ_t0, the correlation diminishes progressively over time, reaching 0.67 six years after publication.

In general, the correlation values are higher, confirming that Mendeley readership exhibits both rapid growth and long-term sustainability from the early stages of the publication lifecycle (Fang, Ho, Han, & Wu, 2024).

In addition to the autocorrelations of impact and readership at different time points, we analyzed all correlations between the time-invariant variables related to publication features (Figure 2).

*Figure 2 - Matrix of correlations among variables*

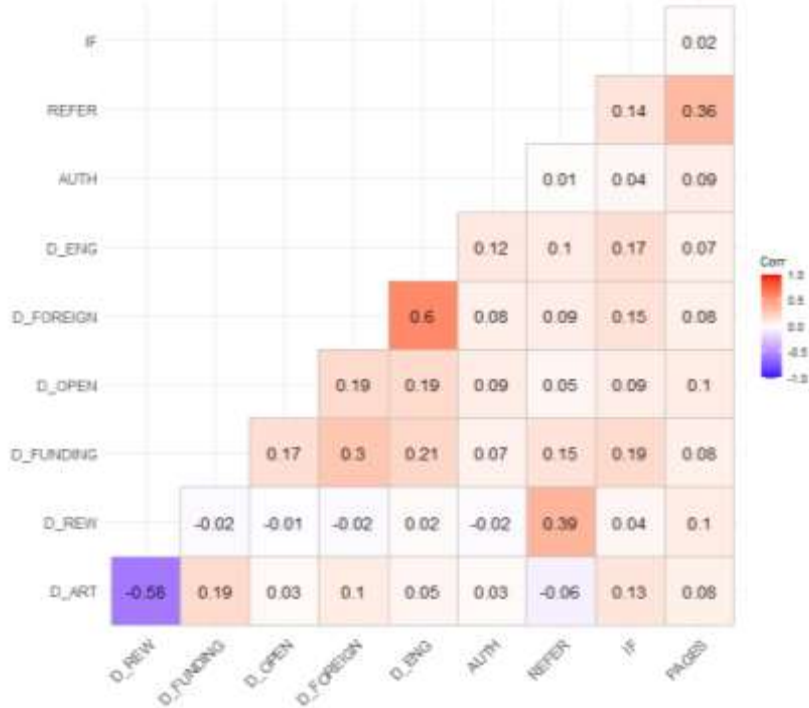

No correlations were sufficiently high to suggest potential multicollinearity in constructing the statistical model, but all are statistically significant. The highest correlation was observed between the dummy variable indicating when the address list contains more than one country (D_FOREIGN) and the dummy variable indicating the presence of a native English-speaking author (D_ENG), with a value of 0.60.

The variable related to the document type 'Review' (REW) is correlated with the number of references (0.39) and inversely correlated with the dummy variable defining publications as articles (ART), as these two categories are mutually exclusive. Lastly, the correlation between the number of references (REFER) and the number of pages (PAGES) is 0.36. No other notable correlations were identified.

**4.2 OLS models**

Table 3 presents the coefficients of a linear regression model across seven iterations ($t_0$-$t_6$) for the "Full" model, offering insights into the evolving significance of various explanatory variables over time. The results show that the coefficient for L_IMPACT_t steadily increases from 0.243 at $t_0$ to 0.976 at $t_6$, indicating that the impact of this variable becomes progressively more significant with each iteration. The constant statistical significance (denoted by ***) across all time points underscores the robustness of its influence throughout the model's development.

Conversely, L_READ_t shows a decreasing trend, starting at 0.278 at $t_0$ and dropping to 0.070 at $t_6$. This suggests a diminishing role of readings over time, although the variable remains statistically significant in every iteration, indicating that it continues to contribute to the model, albeit with less influence as the years go by. Similarly, L_AUTH begins with a positive coefficient at $t_0$ (0.046) but turns negative from $t_2$ onwards, stabilizing at approximately -0.017. This shift suggests that the impact of the number of authors becomes detrimental in later stages, with its influence weakening and eventually becoming counterproductive.

Something similar concerns the variable D_ENG, which starts with a small positive coefficient at $t_0$ (0.035) but turns negative by $t_3$ (-0.001) and further decreases to -0.002 at $t_6$. Initially, the "linguistic advantage" of having English-speaking co-authors in the byline appears to be in place, but this reverses over time. For D_FOREIGN, the coefficient declines from 0.019 at $t_0$ to 0.001 at $t_6$, indicating a reduced influence of the presence of almost another country as iterations progress, though it remains statistically significant.

In contrast, D_FUNDING demonstrates a negative trajectory, with the coefficient turning from 0.010 at $t_0$ to -0.006 at $t_1$ and further declining to -0.001 at $t_6$. This pattern suggests that funding exerts an increasingly negative influence, particularly in the later stages of the model.

D_OPEN shows a gradual decline from 0.035 at $t_0$ to 0.004 at $t_6$, retaining its statistical significance throughout, suggesting that while openness remains a positive contributor, its impact lessens over time. Similarly, L_PAGES exhibits a decreasing coefficient, moving from 0.061 at $t_0$ to 0.016 at $t_6$. Although its influence diminishes, it remains statistically significant, indicating that the number of pages continues to play a role, albeit a smaller one in later iterations.

The behavior of L_IF is noteworthy, starting with a significant positive coefficient at $t_0$ (0.285) but decreasing and eventually turning negative at $t_6$ (-0.008). This suggests that the impact factor (IF) initially contributes positively to the model but loses relevance and becomes a negative factor as iterations progress. The same trend is observed in D_ART, where the coefficient decreases from 0.054 at $t_0$ to 0.010 at $t_6$, reflecting a declining influence of being an article over time, though it remains statistically significant.

Both D_REW and L_REFER exhibit similar behavior. D_REW starts with a strong positive coefficient (0.141 at $t_0$) but turns negative at $t_3$ (-0.005) and continues to decrease to -0.011 at $t_6$. L_REFER follows a comparable path, starting at 0.095 at $t_0$ and turning negative at $t_3$ (-0.001), further declining to -0.006 at $t_6$. These results suggest that the initial positive impact of being a review and the number of references diminishes over time, and by the later iterations, these variables contribute negatively to the model.

The performance of the "Full model" improves significantly across iterations, as demonstrated by the adjusted R², which increases from 0.758 at $t_0$ to 0.975 at $t_6$. This indicates a substantial enhancement in the model's explanatory power as more iterations are considered. Similarly, the F-statistic increases markedly, from 3542.771 at $t_0$ to 43539.748 at $t_6$, signaling a strong improvement in the model's overall significance. Additionally, the residual standard error (RSE) decreases progressively, from 0.363 at $t_0$ to 0.117 at $t_6$, reflecting a reduction in prediction error and an overall improvement in the accuracy of the model. The p-value for the F-statistic remains consistently at 0.000 across all iterations, confirming the statistical significance of the model at each stage.

*Table 3 – Regression coefficients and diagnostic statistics about the Full models, with seven iterations (from t=0 to t=6).*

| Variables | t=0 | t=1 | t=2 | t=3 | t=4 | t=5 | t=6 |
|---|---|---|---|---|---|---|---|
| L_IMPACT_t | 0.243*** | 0.523*** | 0.717*** | 0.833*** | 0.899*** | 0.944*** | 0.976*** |
| L_READ_t | 0.278*** | 0.242*** | 0.187*** | 0.143*** | 0.115*** | 0.091*** | 0.070*** |
| L_AUTH | 0.046*** | 0.004*** | -0.012*** | -0.017*** | -0.018*** | -0.018*** | -0.017*** |
| D_ENG | 0.035*** | 0.011*** | 0.002* | -0.001 | -0.001 | -0.002** | -0.002*** |
| D_FOREIGN | 0.019*** | 0.009*** | 0.005*** | 0.003*** | 0.002*** | 0.001* | 0.001** |
| D_FUNDING | 0.010*** | -0.006*** | -0.009*** | -0.008*** | -0.006*** | -0.004*** | -0.001*** |
| D_OPEN | 0.035*** | 0.017*** | 0.010*** | 0.008*** | 0.006*** | 0.005*** | 0.004*** |
| L_PAGES | 0.061*** | 0.050*** | 0.043*** | 0.034*** | 0.026*** | 0.020*** | 0.016*** |
| L_IF | 0.285*** | 0.122*** | 0.049*** | 0.017*** | 0 | 0.001*** | -0.008*** |
| D_ART | 0.054*** | 0.041*** | 0.023*** | 0.015*** | 0.012*** | 0.010*** | 0.010*** |
| D_REW | 0.141*** | 0.059*** | 0.013*** | -0.005** | -0.010*** | -0.012*** | -0.011*** |
| L_REFER | 0.095*** | 0.039*** | 0.011*** | -0.001 | -0.005*** | -0.006*** | -0.006*** |
| Adjusted R² | 0.758 | 0.846 | 0.900 | 0.931 | 0.951 | 0.965 | 0.975 |
| F-statistics | 3,542.771 | 6,214.632 | 10,219.890 | 15,387.436 | 22,116.089 | 31,240.159 | 43,539.748 |
| p-value F | 0.000 | 0.000 | 0.000 | 0.000 | 0.000 | 0.000 | 0.000 |
| RSE | 0.363 | 0.289 | 0.233 | 0.193 | 0.163 | 0.138 | 0.117 |
| DoF | 293,279 | 293,279 | 293,279 | 293,279 | 293,279 | 293,279 | 293,279 |

*Control variables 247 SC dummies. SC number 1 is considered as the baseline– see supplemental material (SI-Appendix A.xlsx)*

The second table (Table 4) presents results from the Reduced model over seven iterations ($t_0$ - $t_6$), which includes fewer variables compared to the previous model.

In the Reduced model, L_IMPACT_t0 begins at a much higher value than the previous "Full model", 0.529 at $t_0$, and grows to 1.048 at $t_6$. The impact of this variable is stronger from the beginning, and it also grows more quickly, achieving near-maximal values earlier in the iteration process. By $t_4$, the coefficient is already at 1.024, compared to 0.899 in the full model. In other words, L_IMPACT_t0 has a higher and more immediate impact on the outcome, indicating that in the absence of other variables, this factor exerts a more dominant influence earlier in the process.

L_IF follows a similar pattern in both models, but its influence diminishes more rapidly in the Reduced model. Without other variables, the impact factor contributes strongly to the early iterations but quickly loses importance.

*Table 4 – Regression coefficients and diagnostic statistics about the Reduced models, with seven iterations (from t=0 to t=6).*

| Variables | t=0 | t=1 | t=2 | t=3 | t=4 | t=5 | t=6 |
|---|---|---|---|---|---|---|---|
| L_IMPACT_t0 | 0.529*** | 0.790*** | 0.929*** | 0.993*** | 1.024*** | 1.040*** | 1.048*** |
| L_IF | 0.454*** | 0.200*** | 0.086*** | 0.038*** | 0.014*** | 0.003*** | -0.001* |
| Adjusted R² | 0.711 | 0.828 | 0.893 | 0.928 | 0.949 | 0.964 | 0.974 |
| F-statistics | 2,904.625 | 5683.263 | 9,875.971 | 15,240.533 | 22,143.605 | 31,515.997 | 44,179.767 |
| p-value F | 0.000 | 0.000 | 0.000 | 0.000 | 0.000 | 0.000 | 0.000 |
| RSE | 0.396 | 0.305 | 0.240 | 0.197 | 0.166 | 0.140 | 0.119 |
| DoF | 293,289 | 293,289 | 293,289 | 293,289 | 293,289 | 293,289 | 293,289 |

*Control variables 247 SC dummies. SC number 1 is considered as baseline - see supplemental material (SI-Appendix B.xlsx)*

Lastly, Table 5 presents results for the third model, where the regressors are limited to IMPACT only (across different time points from $t_0$ to $t_6$) and SC controls. This "Completely Reduced" model aims to isolate the effects of early citations and disciplinary differences by excluding the control for the impact factor of the hosting source. The coefficient for L_IMPACT at time 0 is very high (0.660) and the differences with the previous models tend to diminish as the time increases. At time 6, no differences are observed in the IMPACT coefficients between this model and the previous Reduced model.

*Table 5 – Regression coefficients and diagnostic statistics about the Completely Reduced models, with seven iterations (from t=0 to t=6).*

| Variables | t=0 | t=1 | t=2 | t=3 | t=4 | t=5 | t=6 |
|---|---|---|---|---|---|---|---|
| L_IMPACT_t0 | 0.660*** | 0.861*** | 0.962*** | 1.007*** | 1.029*** | 1.042*** | 1.048*** |
| Adjusted $R^2$ | 0.665 | 0.820 | 0.892 | 0.928 | 0.949 | 0.964 | 0.974 |
| F-statistics | 2345.407 | 5398.664 | 9769.051 | 15239.953 | 22215.850 | 31641.589 | 44357.504 |
| p-value F | 0.000 | 0.000 | 0.000 | 0.000 | 0.000 | 0.000 | 0.000 |
| RSE | 0.427 | 0.312 | 0.242 | 0.198 | 0.166 | 0.140 | 0.119 |
| DoF | 293290 | 293290 | 293290 | 293290 | 293290 | 293290 | 293290 |

*Control variables 247 SC dummies. SC number 1 is considered as baseline - see supplemental material (SI-Appendix C.xlsx)*

### 4.3 Comparing OLS models

Figure 3 presents a comparative analysis of the three regression models evaluated using Adjusted R-squared (Adjusted $R^2$) across various values of IMPACT_t. Adjusted $R^2$ serves as an indicator of the explanatory power of each model, accounting for the number of predictors included, and thus provides a more robust measure than the standard $R^2$, especially in models with multiple variables.

The comparison shows that the Full model consistently exhibits the highest Adjusted $R^2$ values across the first three iterations, confirming that it has the strongest explanatory capacity among the three models.

*Figure 3 – Comparison of Adjusted $R^2$ statistics after OLS regressions for Full, Reduced models, and Completely Reduced models (for each citation time window t).*

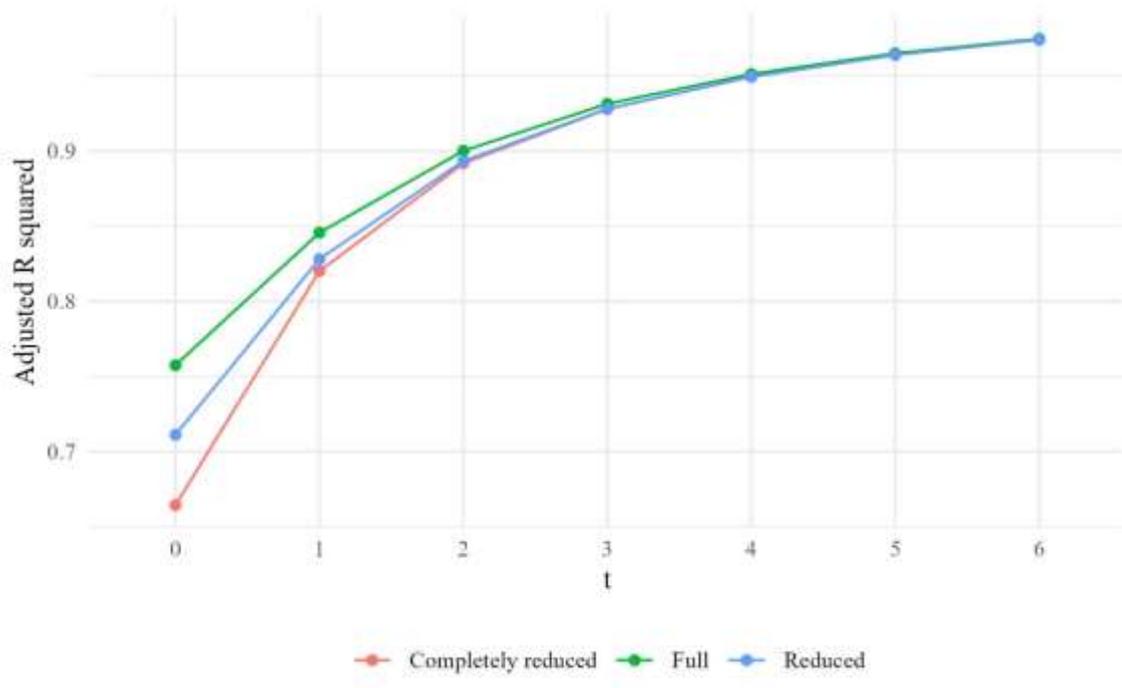

To compare the "Full" model with the "Reduced" model and, subsequently, with the "Completely Reduced" model, fourteen ANOVA tests were conducted, one for each variant of each model comparison. In this context, the ANOVA test evaluates whether the additional predictors included in the more complex models significantly improve their explanatory power. The F-statistic obtained from the ANOVA provides a measure of how much better the Full model explains the variability in the data compared to simpler models, helping to determine if the increase in model complexity is statistically justified. This statistic is given by the following ratio:

$$F = \frac{\left(\dfrac{RSS_{reduced\ (or\ completly\ reduced)} - RSS_{full}}{p}\right)}{\left(\dfrac{RSS_{full}}{n-p-1}\right)}$$

[1]

Where:
- $RSS_{Reduced\ (or\ Completely\ Reduced)}$ is the residual sum of squares of the "Reduced" (or "Completely Reduced") model,
- $RSS_{full}$ is the residual sum of squares of the "full" model,
- p is the number of additional variables in the full model,
- n is the total number of observations.

A high F value indicates that the more complex model explains a greater portion of the variability in the data compared to the Reduced model. All comparisons yielded an F value with a p-value < 0.01, meaning that the "Full" model is consistently more significant overall than the "Reduced" and the "Completely Reduced" one. The F-statistic values are represented in Figure 4 for each model variant.

We can observe that the greater predictive power of the full model decreases with the citation time window, as the contribution of non-scientific factor decreases in favor of that of early citations. In fact, the F-statistic starts at over 5,000 in the ANOVA model comparing the Full and Reduced models and at over 10,000 in the ANOVA model comparing the Full and Completely Reduced models but drops significantly to below 1,000 for both comparisons by $t_6$.

*Figure 4 – Comparison of F values statistics after ANOVA among Full and Reduced models and among Full and Completely Reduced models (for each citation time window t)*

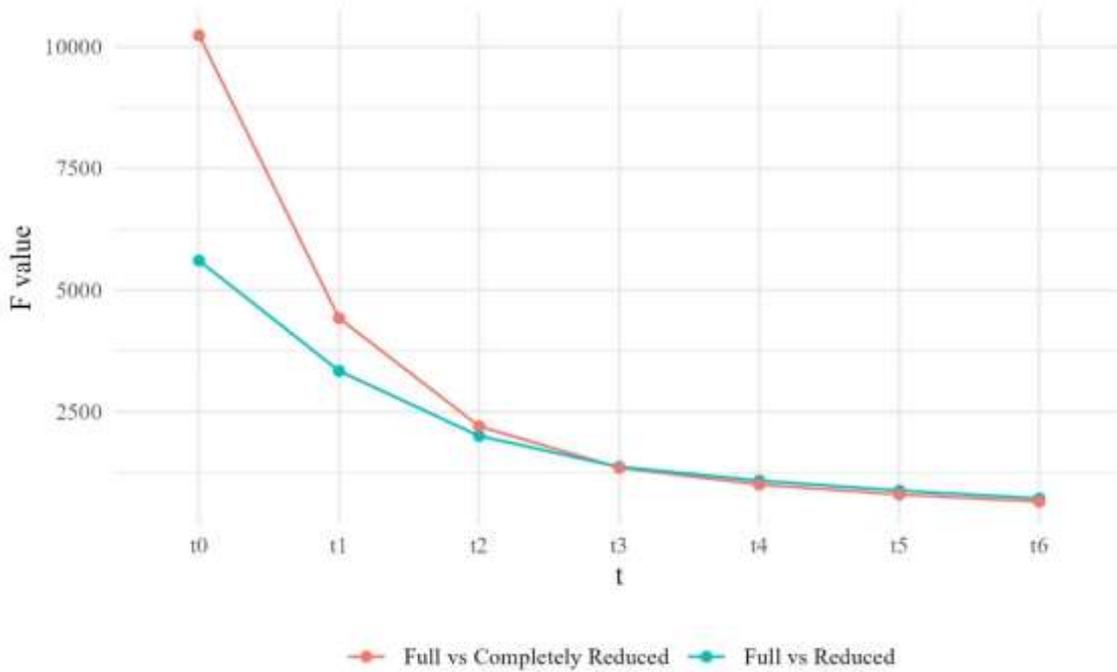

**4.4 Post-estimation analysis**

To assess the adequacy of the regression model, it is crucial to examine the distribution of the residuals and ensure that the assumptions of normality are met. Deviations from normality can affect the validity of hypothesis tests, the precision of confidence intervals, and the overall reliability of the model. Diagnostic plots such as histograms of residuals and Q-Q (Quantile-Quantile) plots are commonly used to evaluate this assumption. These plots provide visual insights into how well the residuals conform to the expected normal distribution.

For example, Figure 5 illustrates the normality check for residuals for the Full model version at $t_0$. The histogram in the left panel displays the distribution of the residuals. It is overlaid with a kernel density estimation (red line), which provides a smoothed estimate of the data distribution. The goal of this plot is to compare the empirical distribution of the residuals to the expected normal distribution. The residuals follow a roughly normal distribution, with most values concentrated around zero. The bell-shaped curve of the kernel density (red line) fits the histogram reasonably well. While the residuals generally exhibit a regular shape, there are slight deviations at the tails, where the distribution does not perfectly align with the normal density. These deviations, though small, could indicate minor departures from normality in the residuals. However, some caution is advised due to the visible deviations, which might impact the model's assumptions in more extreme observations. The Q-Q plot on the right side of Figure 5 is a graphical tool for assessing whether the residuals follow a normal distribution by comparing the quantiles of the sample residuals to the theoretical quantiles from a standard normal distribution. The points largely follow the straight 45-degree reference line in the center of the plot, which indicates that the central part of the residuals conforms well to the normal distribution. However, deviations from the line are visible at the extremes of the distribution, in particular with upper quantiles showing upward curvature, suggesting that the residuals are not perfectly normally distributed. These departures, particularly in the extremes, are indicative of potential outliers or heavy-tailed behavior, which may affect the validity of inferences made using the model.

*Figure 5 – Histogram of residuals and QQ-norm for Full model with impact and readership at time 0*

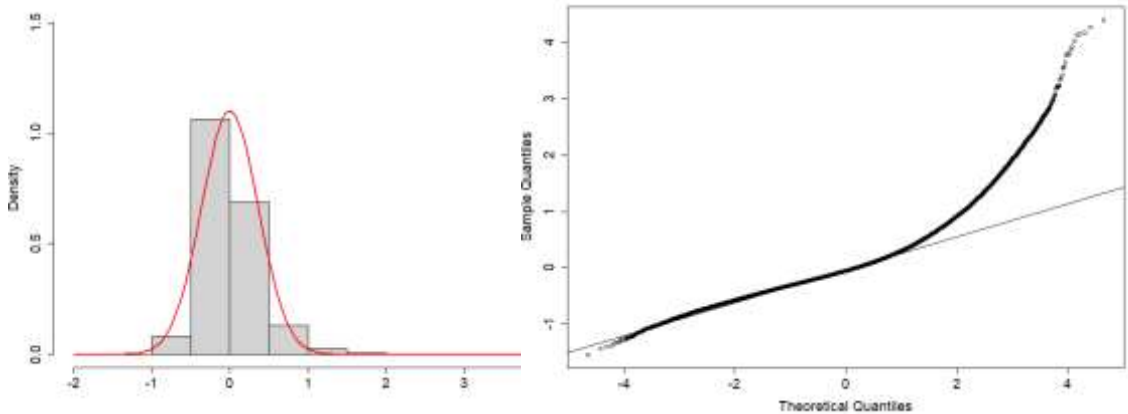

Therefore, we applied a bootstrap procedure. Bootstrapping is a non-parametric technique that allows for the estimation of the distribution of the model coefficients by resampling the observed data. In cases where errors deviate from normality, traditional coefficient estimates can become less reliable, and bootstrapping offers an effective approach to obtain more robust estimates and to assess the variability of the coefficients more accurately. For example, we applied this procedure to the full model with the "impact" variable at time $t_0$. Through the bootstrap method, for each model coefficient, we obtained two values:
- *Bias*, i.e. the difference between the average of the bootstrap estimates and the original estimate, indicating any potential deviation. Bias values near zero suggest that the bootstrap estimates align well with the original estimates.
- *Standard Error*, i.e. the standard deviation of the coefficients obtained from the bootstrap samples, providing an empirical measure of the variability of the estimates.

As evidenced in the supplemental material (SI-Bootstrap_results.xlsx), the values of *Bias* and *Standard Error* obtained for 500 iterations/samples are minimal and thus acceptable, confirming the stability of the estimates and the validity of the model.

Finally, in Figure 6, we show the distribution of coefficients recorded for the 247 dummies related to as many SCs for each citation time window. The temporal pattern is evident, with frequency distributions tending to shift to higher values as the citation time window increases. For $t_0$, the coefficients are all invariably negative and the mean is -0.4, for $t_1$ the coefficient is positive for 6.5% of all 247 SCs and the average increase to -0.175. For $t_6$, the coefficient is positive for 42.5% of SCs, and the average becomes nihil. The variability of the distributions (evidenced by the shown interquartile ranges) tends to decrease with the citation time window.

*Figure 6 –Box plot of coefficients of SC dummies for the FULL model, by citation time window*

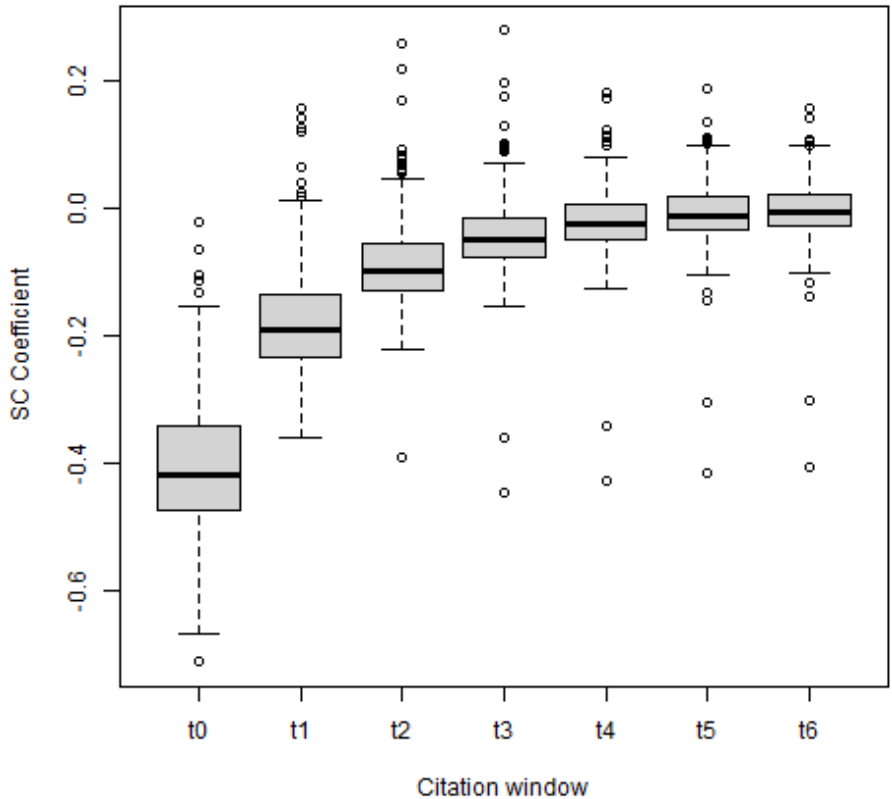

## 5. Discussion and conclusions

This study sought to enhance the prediction of long-term scholarly impact by integrating early citation data with Mendeley readership metrics and various non-scientific factors, such as journal impact factor, authorship and reference list characteristics, and publication accessibility. Our findings reveal that early citation trends and readership data, when combined with these additional attributes, yield significantly improved predictions of a publication's long-term influence, thereby advancing models that traditionally rely on citations alone.

Our findings confirm that early citations, particularly within the first few years post-publication, remain among the strongest indicators of future impact. This result is consistent with earlier studies (Bornmann & Daniel, 2010), reinforcing the predictive value of early scholarly attention. However, by incorporating additional predictors—specifically Mendeley readership, journal impact factor, and characteristics of authorship and reference lists—the proposed model consistently outperforms prior models that include only early citations and IF (Abramo et al., 2019), as well as the widely used model based on early citations alone. This is especially evident in citation windows from 0 to 2 years, where the enhanced model achieves more precise predictions of long-term impact. These results suggest that while early citations are invaluable, they alone cannot fully capture the complex factors contributing to a publication's sustained influence.

Among the non-scientific factors, readership metrics on Mendeley proved particularly valuable. Early readership correlates strongly with subsequent citations, suggesting that readership captures a publication's broader visibility and engagement beyond immediate scholarly circles. Journal impact factor and authorship characteristics, including international collaboration and native English-speaking authorship, also emerged as significant predictors, underscoring the importance of the publication venue and the social-linguistic context of authorship.

Interestingly, while initially associated with enhanced visibility and early citations, open access status and funding showed diminishing effects over time. This trend suggests that while open access and funding contribute to initial visibility, their long-term effect on citations may wane as the novelty of the work decreases or as the field evolves. Such dynamics underscore the importance of time-sensitive models that account for changes in impact predictors across the life cycle of a publication.

Theoretically, this study contributes to the understanding of scientific impact prediction by validating that integrating non-citation-based factors with early citation counts provides a more holistic and accurate prediction of a publication's long-term influence. Traditional citation-based models focus on scholarly attention alone, potentially overlooking non-scholarly engagement and visibility metrics that offer additional insight into a publication's reach and significance. By including non-scientific attributes, we offer a model that better reflects the complex, multifaceted drivers of research impact, highlighting how both immediate scholarly responses and broader external factors shape a work's long-term influence.

Practically, these findings have important implications for stakeholders such as research funders, institutions, and individual researchers. For funding bodies and policymakers, the enhanced model offers a refined tool for evaluating early indicators of high-impact publications, allowing for more informed decisions about funding allocations and research priorities. For institutions, this model underscores the value of encouraging open-access publishing, supporting collaborative research across countries, and pursuing publication in high-impact journals—all factors contributing to a publication's long-term success. Furthermore, for researchers, the findings suggest strategies for increasing the likelihood of long-term impacts, such as targeting high-visibility journals, engaging in international collaborations, and considering open-access options to maximize the work's reach and accessibility.

In addition, this study's findings open potential pathways for enhancing research evaluation practices in institutional and funding agency settings. By adopting a more comprehensive model that includes both citation and non-citation factors, evaluators can assess emerging research with greater precision, capturing the broader spectrum of factors that predict influence beyond academia alone. This approach could lead to a more balanced evaluation system that values not only scholarly impact but also public and interdisciplinary engagement.

While our study advances impact prediction methods, it is important to acknowledge certain limitations. First, our analysis focused on publications from Italy within a specific timeframe (2010–2012), which may limit the generalizability of the findings to other contexts. It is plausible that the importance of specific predictors varies across fields and countries due to differences in publication practices, collaboration networks, and access to funding. Future research could extend this work by testing the model on data from diverse regions, publication years, and scientific disciplines to assess the robustness of these predictors.

Second, although Mendeley readership and citation metrics are strong indicators of academic engagement, their predictive power could potentially be improved by incorporating broader altmetrics. Social media platforms, news mentions, and policy document citations may capture additional layers of public and interdisciplinary interest not reflected in traditional citation data alone. As digital media increasingly shapes the dissemination of knowledge, future studies could examine how these alternative metrics contribute to long-term impact, especially in fields that engage heavily with non-academic audiences.

Another limitation involves the focus on quantitative predictors alone. Non-numeric characteristics, such as the novelty of research ideas, the clarity of communication, or the interdisciplinary nature of the work, are likely influential but challenging to quantify. Machine learning approaches, such as natural language processing (NLP) techniques, could be explored to assess text features like abstract complexity, thematic novelty, or interdisciplinarity. By integrating these content-driven insights, future models might offer an even more nuanced prediction of long-term influence.

In summary, our study demonstrates that combining early citation metrics with a range of non-citation-based factors results in more accurate predictions of a publication's long-term impact. This approach not only expands traditional models but also offers a more nuanced understanding of how publications achieve influence over time. By capturing both scholarly and non-scholarly dimensions of impact, our model provides a practical, accessible framework for assessing research influence in its early stages, empowering funders, institutions, and researchers to make better-informed decisions about the potential value of emerging research.

As the scholarly landscape continues to evolve, this integrated model represents a promising step toward more timely, comprehensive research evaluation tools that align with the multifaceted nature of scientific impact.